\documentclass[10pt,letterpaper,twocolumn]{article} 

\newcommand{\dblo}[1]{\overline{\overline{#1}}}

\usepackage{ol2SmallCaptions}
\usepackage{comment}
\usepackage[draft]{hyperref}
\usepackage{amsmath}
\usepackage{subfigure}    
\usepackage{empheq}  
\usepackage{graphicx}	
\usepackage{epstopdf}

\begin{document}

\twocolumn[ 

\title{Dark state lasers}  \vspace{-10pt}


\author{Cale M. Gentry$^*$ and Milo\v{s} A. Popovi\'{c}$^\dagger$}
\address{
Department of Electrical, Computer, and Energy Engineering, University of Colorado Boulder, Colorado, 80309-0425, USA\\
$^*$cale.gentry@colorado.edu, $^\dagger$milos.popovic@colorado.edu
}\vspace{-10pt}

\begin{abstract}
\noindent We propose a new type of laser resonator based on imaginary ``energy-level splitting'' (imaginary coupling, or quality factor Q splitting) in a pair of coupled microcavities.  A particularly advantageous arrangement involves two microring cavities with different free-spectral ranges (FSRs) in a configuration wherein they are coupled by ``far-field'' interference in a shared radiation channel.  A novel Vernier-like effect for laser resonators is designed where only one longitudinal resonant mode has a lower loss than the small signal gain and can achieve lasing while all other modes are suppressed.  This configuration enables ultra-widely tunable single-frequency lasers based on either homogeneously or inhomogeneously broadened gain media.  The concept is an alternative to the common external cavity configurations for achieving tunable single-mode operation in a laser.  The proposed laser concept builds on a high-Q ``dark state'' that is established by radiative interference coupling and bears a direct analogy to parity-time (${\cal{PT}}$) symmetric Hamiltonians in optical systems.  Variants of this concept should be extendable to parametric-gain based oscillators, enabling use of ultrabroadband parametric gain for widely tunable single-frequency light sources.
\end{abstract}

\vspace{-10pt}

 ] 
\noindent Lasers traditionally comprise a gain medium and a resonant cavity \cite{Verdyn95}.  The resonant cavity has a free spectral range (FSR) inverse to its round-trip length.  If the gain bandwidth spans many FSRs, as often does in free space lasers, multiple longitudinal modes of the cavity see net gain resulting in mode competition.  With inhomogeneously broadened gain media all of the lines with net gain lase, while with homogeneously broadened gain media there is mode competition and the one with the highest gain lases.  In the latter case, if a tunable laser is desired, tuning the cavity length can shift the lasing line across the wavelength spectrum.  However, since the gain spectrum is stationary, if a competing line becomes lowest loss during the wavelength tuning, it begins to lase instead, leading to what is known as mode hopping, and requiring an intra-cavity filter (grating) in external cavity tunable lasers (e.g. Littman-Metcalf \cite{Littman} or Littrow \cite{Littrow} configurations).  In general, competing modes are a recurring problem in laser cavities.  Although on-chip lasers implemented in integrated photonics can have very small resonant cavities, it is still conceivable to have gain media that span multiple FSRs of the cavity.  In addition, parametric gain due to $\chi^{(3)}$ nonlinearity is an extremely broadband form of gain if phase matching is properly engineered \cite{lin07}.
\begin{figure}[b]
  \centering
  \vspace{-20pt}
  \includegraphics[width= \columnwidth]{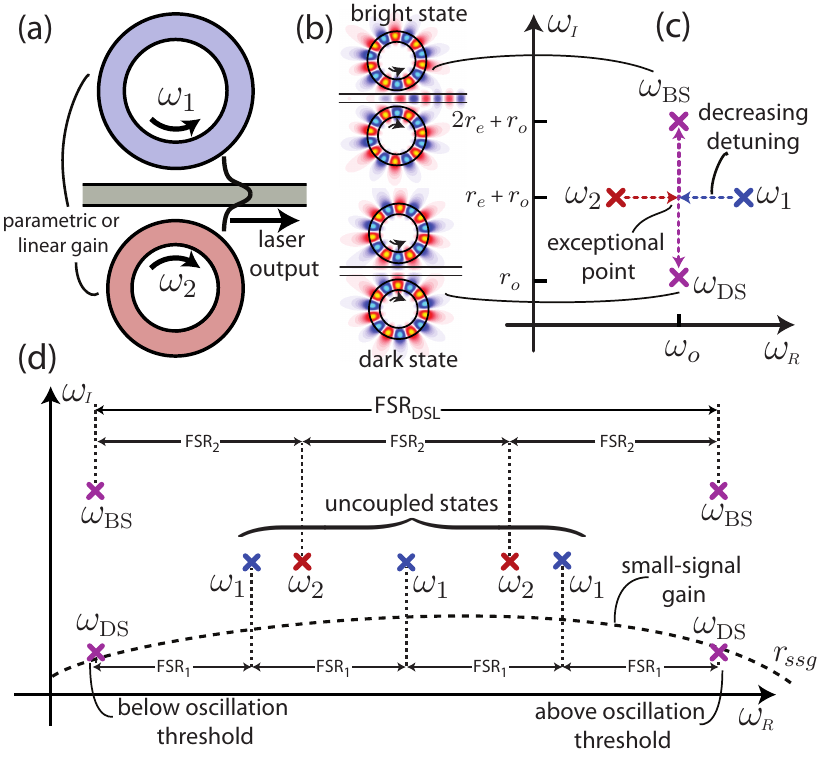}
  \vspace{-25pt}
\caption{(a) Proposed laser resonator geometry enabling far-field interference at the output coupler resulting in (b) a low-Q (bright) state and high-Q (dark) state with (c) eigenfrequency imaginary splitting at matched resonances to create (d) a broad Vernier-like FSR for ultra-wide tuning.\label{fig:sm}}
\end{figure}

In this paper, we propose a fundamentally new type of laser resonator that employs a form of Vernier effect to achieve only a single lasing mode across a very wide tuning range (much greater than the FSR of the cavity) even with arbitrarily broadband gain media.  The laser resonator is based on two resonant cavities which share an output coupler with different FSRs [Fig.~\ref{fig:sm}(a)].  When two resonances are matched, a resonator mode exists that couples to a radiation channel from two cavities with equal amplitude but 180 degrees out of phase.  Destructive interference in the radiation channel leads to a much lower external coupling (higher external Q) than would otherwise be provided by the output coupler to either one of the two cavities in isolation.  On the other hand, at all other resonances the cavities are detuned due to different FSRs and have a low external Q (i.e. strong external coupling).  Hence, only one longitudinal mode has a high enough Q to reach the lasing threshold.

The Vernier effect is well known as a technique to extend the FSR of bandpass filters \cite{Oda91}, and also relies on at least two cavities with differing FSRs.  The type of Vernier effect we harness here is in some ways opposite to that employed in filters, the latter not being usable for a laser cavity.  A filter Vernier effect produces a high-order (e.g. maximally flat) passband where the resonances are matched (this passband is low Q) and a suppressed drop port transmission at mismatched resonances \cite{Oda91}.  Were we to use only one waveguide coupled to the first of a pair of coupled cavities, like a filter without a drop port, all resonances of the second cavity (that is not directly coupled to the waveguide) except the one in the passband would be detuned from the first cavity and hence high Q.  Therefore, if the cavities had gain, all resonances of the second cavity \emph{except} the one with the flat-top bandpass response would be in a position to lase.  In this work, on the other hand, we achieve a single high-Q resonance at the matched frequency, and low Q's at all others.  The key difference is that conventional Vernier filters use traditional (what we'll call ``reactive'') resonator coupling which leads to real frequency splitting, and we use far-field radiative coupling which leads to imaginary splitting, an effect previously proposed and demonstrated \cite{Dahlem10}.  The imaginary splitting leads to a dark state similar but not entirely analogous to those in electromagnetically induced transparency (EIT) \cite{Xu06}.  Because it is the dark state that has the high enough Q to lase, we call the proposed lasers `dark state lasers'.

The basic concept is illustrated in Fig.~\ref{fig:sm}.  We consider two resonant cavities with equal gain, coupled in the far field by an imaginary coupling coefficient. A coupling of modes in time (CMT) model \cite{Haus91,Suh04} describes all of the relevant physics:
$\vspace{-15pt}$

%

\begin{align}
{d \over{dt}} \vec{a} &= j \bar{\bar{\omega}} \cdot \vec{a} - j \bar{\bar{\mu}} \cdot \vec{a} -j \dblo{M}_i s_+ \label{eqn:cmt1}\\
s_- &= -j \overline{\overline{M}}_o \cdot \vec{a} + s_+
\end{align}
\vspace{-15pt}
where
\begin{align*}
\begin{array}{cc}
\vec{a} = \begin{pmatrix}a_1 \\  a_2 \end{pmatrix} \qquad  \qquad \bar{\bar{\mu}} = -j \begin{pmatrix}
r_{\mathrm{e}} & r_{\mathrm{e}} \\
r_{\mathrm{e}} & r_{\mathrm{e}}
\end{pmatrix} \vspace{5pt} \nonumber \\
\dblo{M}_i = \sqrt{2 r_e}\begin{pmatrix}1\\1\end{pmatrix} \qquad \qquad \dblo{M}_o = \dblo{M}_i^T\\
\bar{\bar{\omega}}= \begin{pmatrix}
\omega_{\mathrm{o}} + \delta \omega_{\mathrm{o}} + j(r_{\mathrm{o}} - r_{\mathrm{g}}) & 0 \\
0 & \omega_{\mathrm{o}} - \delta \omega_{\mathrm{o}} + j(r_{\mathrm{o}} - r_{\mathrm{g}})
\end{pmatrix} 
\end{array}
\end{align*}
Here $a_1$ and $a_2$ are the energy amplitudes of the resonant modes in the two cavities; $\omega_{\mathrm{o}} \pm \delta \omega_{\mathrm{o}}$ are the individual, uncoupled resonance frequencies of the two cavities whose detuning $2\delta \omega_{\mathrm{o}}$ can be controlled; and $r_{\mathrm{e}}$, $r_{\mathrm{o}}$, and $r_{\mathrm{g}}$ are the decay rate due to external coupling, decay rate due to intrinsic loss (radiation loss, roughness loss, absorption), and the gain rate due to an optical gain, respectively. We have assumed (without loss of generality) that the decay/gain rates are the same in each cavity and that there is negligible direct (real) coupling between the cavities.
Solving the system (\ref{eqn:cmt1}) in the steady state for zero input ($s_+ = 0$) gives the eigenfrequencies (resonances) of the system
\vspace{-10pt} 
\begin{align}
\omega_\pm = \omega_{\mathrm{o}} + j(r_{\mathrm{o}}+r_{\mathrm{e}}-r_{\mathrm{g}} \pm \sqrt{r_{\mathrm{e}}^2-\delta \omega_{\mathrm{o}}^2})
\label{eqn:eigfreq}
\end{align}
with corresponding eigenvectors (supermodes) of the form 
\vspace{-10pt}
\begin{align}\label{eq:supermodes}
\begin{pmatrix} a_1 \\ a_2 \end{pmatrix}_\pm = {1\over{C}}\begin{pmatrix} 1 \\ \pm \sqrt{1- ({\delta \omega_{\mathrm{o}} \over{r_{\mathrm{e}}}})^2} + j {\delta \omega_{\mathrm{o}} \over{r_{\mathrm{e}}}} \end{pmatrix}
\end{align}
where C is a normalization constant.

We next explore the salient features of this system.  In the range of detunings smaller than the external coupling (i.e. $\delta \omega_{\mathrm{o}} < r_{\mathrm{e}}$), it is evident that both supermodes have equal \emph{real} resonant frequencies at the arithmetic mean of the individual resonator cavities' uncoupled resonant frequencies. Therefore, the imaginary coupling term results in an ``attraction'' of resonant frequencies, illustrated in Fig.~\ref{fig:sm}(c), in contrast to the usual energy level repulsion that is prototypical of a real (reactive) coupling term \cite{Haus91}.  Instead of splitting along the real frequency axis the eigenfrequencies split along the imaginary axis.  Physically this means there is no energy exchange coupling between the individual cavities while the corresponding quality factors, $Q = {Re\{\omega\}\over{2Im\{\omega\}}}$ \cite{Vahalla04}, for the two supermodes split due to interaction at the point of coupling to the shared radiation channel. Q-splitting has been demonstrated via far-field interference in radiation loss \cite{Vahalla04,Benyoucef11} and in a single-mode external coupling bus radiation channel \cite{Dahlem10, Xu06}.  It has also been demonstrated in Bragg matching of momentum detuned modes to achieve scatterer-avoiding cavity supermodes \cite{ShainlineCLEOwigglers,YangyangIPR,ShainlineOLwigglermodulators}. Imaginary k-splitting, a waveguide equivalent to Q-splitting, has also been demonstrated for ultra-low-loss waveguide crossings \cite{YangyangOL}. 
\begin{figure}[t]
  \centering
  \includegraphics[width= \columnwidth]{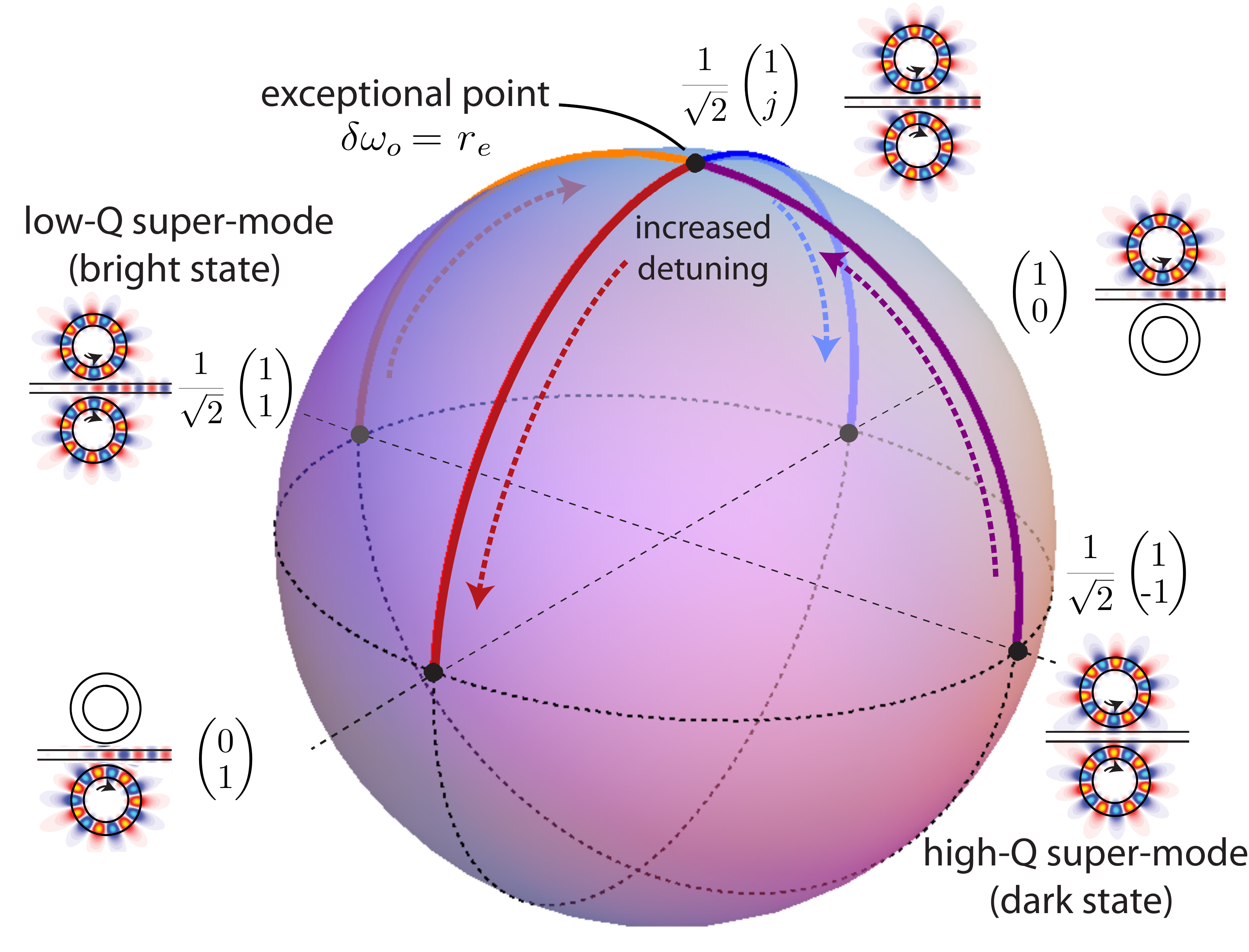}
\vspace{-25pt}
\caption{Visualization of supermodes with detuning on a Poincar\'{e} (Bloch) sphere. For detuning less than the single ring external coupling ($\delta \omega_{\mathrm{o}} < r_{\mathrm{e}}$) the energy is equally distributed across both rings. At greater detuning the supermodes approach the modes of the individual uncoupled rings. \label{fig:ps}}
\vspace{-25pt}
\end{figure}


As a laser cavity structure we propose two ring resonators with different FSRs coupled to a waveguide as shown in Fig.~\ref{fig:sm}(a), generalizing the geometry first proposed and demonstrated as a passive slow light cell \cite{Dahlem10}. Similar structures have been proposed for loadable and erasable optical memory units \cite{Ding08}, whose transmission characteristics have been theoretically investigated via both CMT \cite{Zhang12} and transfer matrix method \cite{Zhang07, Zhang11}. The supermodes in the case of zero detuning are  illustrated in Fig.~\ref{fig:sm}(b) and consist of a high-loss `bright state', $\vec{a}_{\mathrm{BS}} = {1\over{\sqrt{2}}} (1,1)^T$ at frequency $\omega_{\mathrm{BS}}$, with large external coupling and an antisymmetric, low-loss `dark state', $\vec{a}_{\mathrm{DS}} ={1\over{\sqrt{2}}} (1,-1)^T$ at frequency $\omega_{\mathrm{DS}}$, with zero external coupling. The corresponding resonant frequencies are split along the imaginary axis, $\omega_{\mathrm{BS}}-\omega_{\mathrm{DS}}= j 2 r_e$. Therefore, for a small-signal gain, described by gain rate $r_\mathrm{ssg}$, larger than the intrinsic loss rate $r_o$, but smaller than the loaded passive decay rate $r_{\mathrm{o}}+r_{\mathrm{e}}$, \emph{only} the dark state will be above lasing threshold.  For small signal gain above $r_e+r_o$, resonances at other FSRs may see net gain and begin to lase, in our current discussion an undesirable feature.  If the two cavities have different FSRs then the Q-splitting will only occur where the resonant frequencies match, resulting in an effective FSR between dark states determined by the least common multiple of the FSRs of the individual cavities as illustrated in Fig.~\ref{fig:sm}(d). This allows for a Vernier-like selection (and tuning) effect over an ultra-wide wavelength range.

\begin{figure}[t!]
  \centering
  \includegraphics[width= \columnwidth]{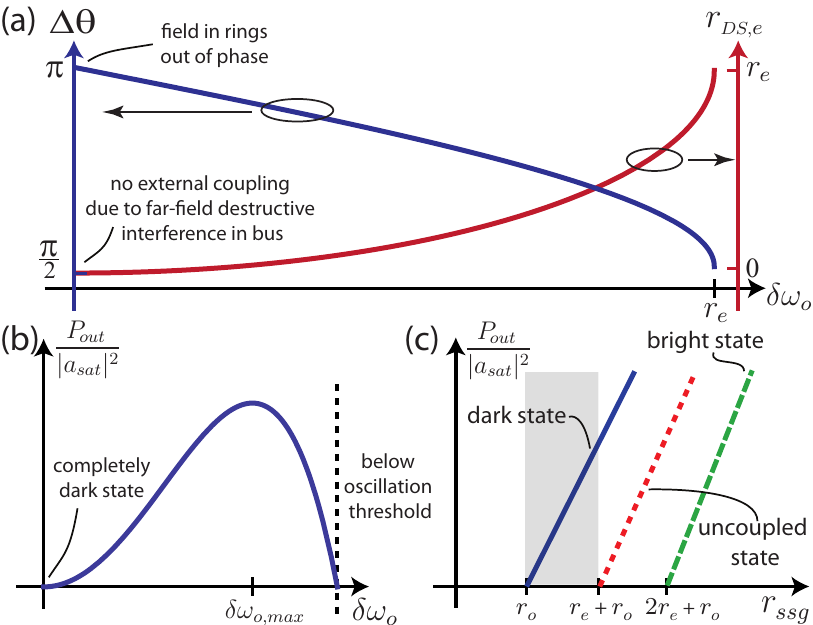}
  \vspace{-25pt}
\caption{(a)Phase difference between mode-fields of each ring and resulting external coupling of dark state with detuning. (b)Dependence of tuning on output power for fixed $r_{\mathrm{e}}$ (c)Threshold/slope efficiency curves as a function of small-signal gain (pumping), with the operating region of dark-state-only lasing shaded. \label{fig:threshold}}
\vspace{-15pt}
\end{figure}

The dark state is named as such because there is exactly no coupling of cavity light energy into the output waveguide.  As with any laser, for dark state lasing to be useful there must be a finite external output coupling. This is achieved via a slight detuning of the resonators. The dependence of laser output on detuning is illustrated in Fig.~\ref{fig:threshold}(b). Since the supermodes are, in general, two-dimensional complex eigenvectors normalized to unit energy they can be visualized similarly to polarization (spin) on a Poincar\'{e} (Bloch) sphere. Fig.~\ref{fig:ps} illustrates the evolution of the supermodes with increased detuning. The dependence of the total external coupling of the dark state supermode on detuning is described by
\begin{align}
r_{\mathrm{DS, e}} = r_{\mathrm{e}} - \sqrt{r_{\mathrm{e}}^2-\delta\omega_{\mathrm{o}}^2}
\vspace{-5pt}
\end{align}
Physically, this finite external coupling results from the no longer perfect destructive interference in the waveguide due to the phase difference between the cavities deviating from ${\pi}$ with detuning [Fig.~\ref{fig:threshold}(a)]. 
This results in a lasing threshold condition on the dark state of $r_{\mathrm{ssg}} > r_{\mathrm{o}}+r_{\mathrm{DS, e}}$. Too high an external coupling will result in the laser dropping below threshold as shown in Fig.~\ref{fig:threshold}(b) at large cavity-cavity detuning $\delta\omega_o$.

To investigate design of the cavity for optimal lasing characteristics we introduce a saturable gain into the model. Assuming equal gain properties in each ring, when the dark state is over threshold the gain rate is
\vspace{-5pt}
\begin{align}
r_{\mathrm{g}} = {r_{\mathrm{ssg}}\over{1 + {|a_{\mathrm{DS}}|^2\over{|a_{\mathrm{sat}}|^2}}}}.
\vspace{-5pt}
\end{align}
The steady state output of the resonator with a saturable gain medium is to be maximized next.  The output power relative to the saturation energy, ${P_{\mathrm{out}} \over{|a_{\mathrm{sat}}|^2}}$, is maximized at a particular, optimal choice external coupling,
\vspace{-5pt}
\begin{align}
r_{\mathrm{max,e}} = \sqrt{r_{\mathrm{ssg}} r_{\mathrm{o}}}-r_{\mathrm{o}}.
\end{align}

We can consider the saturation of the whole 2-cavity resonator rather than that of each ring individually because in the range of detuning $\delta\omega_o$ where Q-splitting occurs the energy in each ring is equal  (\ref{eq:supermodes}). Here the size difference of the rings is assumed to be negligible with respect to its saturation properties.  From this model, the threshold and slope efficiencies are found to be described by 
\vspace{-5pt}
\begin{align}
{P_{\mathrm{out}} \over{|a_{\mathrm{sat}}|^2}} \!=\!\! 
 \begin{cases} \! \qquad \qquad \quad 0 \qquad \qquad  \hspace{10.5 pt}, \!&\!\!\! r_{\mathrm{ssg}}\!<\!r_{\mathrm{DS,e}}\!+\!r_{\mathrm{o}} \\ \! {2 r_{\mathrm{DS, e}}\over{r_{\mathrm{o}}+r_{\mathrm{DS,e}}}}(r_{\mathrm{ssg}}\!-\!r_{\mathrm{DS,e}}\!-\!r_{\mathrm{o}}), \!&\!\!\! r_{\mathrm{ssg}}\!\geq\!r_{\mathrm{DS,e}}\!+\!r_{\mathrm{o}}  \end{cases} 
\end{align} 
This expression is general to any laser cavity. To find similar parameters for the uncoupled states and bright state $r_{\mathrm{DS,e}}$ is simply replaced with their respective output couplings.  The laser mode outputs as a function of small signal gain are illustrated in Fig.~\ref{fig:threshold}(c).  Note that a high output coupling increases the threshold requirement of a lasing mode but also results in a higher slope efficiency with respect to the small signal gain $r_{\mathrm{ssg}}$.

\begin{figure}[t]
  \centering
  \includegraphics[width= \columnwidth]{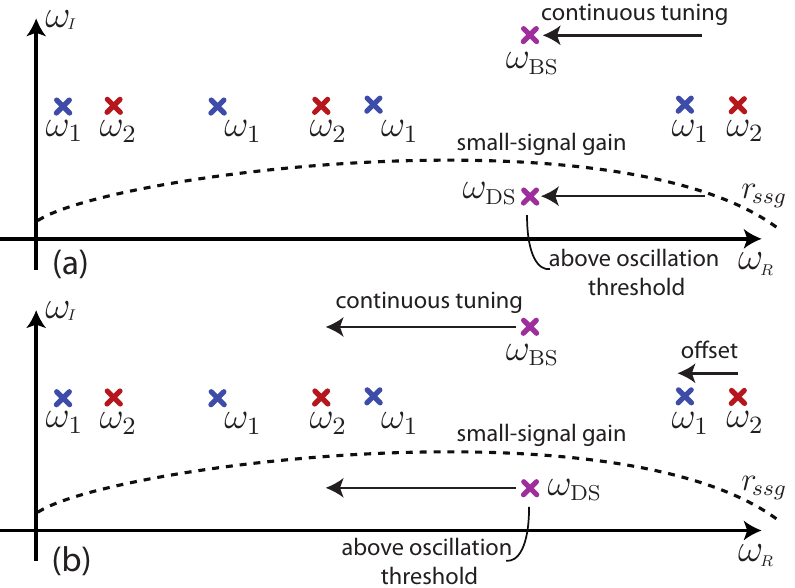}
 \vspace{-25pt}
\caption{An illustration of quasi-continuous tuning of the dark state laser. First (a) both rings are tuned an FSR of the smaller ring (larger FSR). Then the laser can be turned off and reset to the start position Fig 1 (d). (b) An offset can then be introduced to shift the Q-splitting to the previous location where both rings can then be continuously tuned across another FSR. The process can then be repeated across the gain bandwidth. \label{fig:tuning}}
\vspace{-20pt}
\end{figure}

Vernier-like tuning has been used in lasers featuring sampled grating distributed Bragg reflectors with different periods \cite{Jayaraman93}. We note that this is a fundamentally different mechanism than that presented in this paper, not based on imaginary coupling due to far-field interference.  Here, we will briefly outline possible tuning strategies for dark state lasers. Tuning only one of the two rings is the simplest method but results in discontinuous tuning. In this method the resonance frequency of a single ring is shifted resulting in the Q-splitting occurring at a different FSR and therefore shifting at which wavelength the laser is operating. This will result in successful tuning of the laser across a gain bandwidth, albeit in discrete steps of the FSR of the larger ring.  In principle, it is also possible to tune the laser almost continuously over many FSRs if one is able to carefully tune both rings. This can be achieved even if the tuning range of each resonator is limited to less than two FSRs (as it is often difficult to thermally tune across several FSRs). This quasi-continuous tuning strategy is illustrated in Fig.~\ref{fig:tuning}. Both rings can be tuned an FSR of the smaller cavity (larger FSR) where the laser can be briefly shut off and tuned back to the start position in Fig.~\ref{fig:sm}(d). Then one can introduce an offset by detuning the rings until the Q-splitting occurs at the wavelength the laser was operating previously, but at the next longitudinal-order mode pair. This process can continue allowing a quasi-continuous tuning across the dark state FSR.

An interesting property of the dark state laser cavity is the manifestation of an exceptional point at $\delta \omega_{\mathrm{o}} = r_\mathrm{e}$.  The exceptional point is characterized by the coalescing of both eigenvalues and eigenvectors of the system as shown in Fig.~\ref{fig:sm}(c) and Fig.~\ref{fig:ps} respectively, along with a vanishing norm \cite{Heiss04}. This results from a square root branch point in (\ref{eqn:eigfreq}) and physically occurs at the transition from resonant frequency attraction to Q-splitting. The presence of an exceptional point leads to analogies to parity-time (${\cal{PT}}$) symmetric Hamiltonians in quantum mechanics \cite{Bender98}. Recently ${\cal{PT}}$-symmetry analogies in optical systems have been investigated in waveguides \cite{Ruschhaupt05, Guo09} and resonators \cite{Yoo11} with symmetric real refractive index and antisymmetric imaginary refractive index. The dark state laser geometry provides a ``quasi''-${\cal{PT}}$-symmetry breaking at an exceptional point where external coupling takes the place of absorption and the complex resonant frequencies are shifted up along the imaginary axis in the complex frequency plane by the constant loss $r_{\mathrm{o}}$. This is of significance since, unlike in ${\cal{PT}}$-symmetry, in our system no absorption or gain is required for this property to arise.  The dark state system is ${\cal{PT}}$-like after a gauge transformation \cite{Guo09} and under the approximation where coupling to radiation modes is irreversible. Furthermore, the ${\cal{PT}}$-symmetry breaking threshold is lowered to zero in the case of a system with degenerate cavities, $\delta\omega_o = 0$. 

The proposed resonator may enable a new approach to the design of widely tunable laser sources, and in principle extends to table top resonators with a shared output coupler.  The design need not use different size resonators if the Vernier property is not of importance.  More generally, the concepts of imaginary frequency splitting and energy level attraction, quasi-${\cal{PT}}$-symmetric optical systems, and thresholdless quasi-${\cal{PT}}$-symmetry breaking may find applications in other photonic device technology.
\vspace{2pt}

This work was supported by the Packard Foundation and a CU-NIST Measurement Science and Engineering (MSE) Fellowship.


\end{document}